\documentclass[fleqn,usenatbib]{mnras}

\usepackage{newtxtext,newtxmath}
\usepackage[]{amsmath}

\usepackage[T1]{fontenc}
\usepackage{ae,aecompl}

\usepackage{graphicx}	
\usepackage{amsmath}	
\usepackage{amssymb}	

\newcommand{\sm}{\small}
\newcommand{\zobov}{{\small ZOBOV} }

\title[Reconstruction for void RSD]{A Zeldovich reconstruction method for measuring redshift space distortions using cosmic voids}

\author[S. Nadathur et al.]{
Seshadri Nadathur,$^{1}$\thanks{E-mail: seshadri.nadathur@port.ac.uk}
Paul Carter$^{1}$
and Will J. Percival$^{2,3,1}$
\\
$^{1}$Institute of Cosmology and Gravitation, University of Portsmouth, Burnaby Road, Portsmouth PO1 3FX, UK\\
$^{2}$Department of Physics and Astronomy, University of Waterloo, 200 University Ave W, Waterloo, ON N2L 3G1, Canada\\
$^{3}$Perimeter Institute for Theoretical Physics, 31 Caroline St. North, Waterloo, ON N2L 2Y5, Canada
}

\date{Accepted XXX. Received YYY; in original form ZZZ}

\pubyear{2018}

\begin{document}
\label{firstpage}
\pagerange{\pageref{firstpage}--\pageref{lastpage}}
\maketitle

\begin{abstract}
Redshift space distortions (RSD) in the void-galaxy correlation $\xi^s$ provide information on the linear growth rate of structure in low density environments. Accurate modelling of these RSD effects can also allow the use of voids in competitive Alcock-Paczynski measurements. Linear theory models of $\xi^s$ are able to provide extremely good descriptions of simulation data on all scales provided the real space void positions are known. However, by reference to simulation data we demonstrate the failure of the assumptions implicit in current models of $\xi^s$ for voids identified directly in redshift space, as would be simplest using real observational data. To overcome this problem we instead propose using a density-field reconstruction method based on the Zeldovich approximation to recover the real space void positions from redshift space data. We show that this recovers the excellent agreement between theory and data for $\xi^s$. Performing the reconstruction requires an input cosmological model so, to be self-consistent, we have to perform reconstruction for every model to be tested. We apply this method to mock galaxy and void catalogues in the Big MultiDark $N$-body simulation and consistently recover the fiducial growth rate to a precision of $3.4\%$ using the simulation volume of $(2.5\;h^{-1}\mathrm{Gpc})^3$.
\end{abstract}


\section{Introduction}
\label{sec:introduction}
Galaxy redshift surveys provide three-dimensional redshift-space maps of the large-scale structure of the Universe that contain anisotropies along the line of sight direction thanks to two effects: redshift space distortions (RSD) \citep{Kaiser:1987} due to gravitationally-induced peculiar velocities, and the Alcock-Paczynski (AP) effect \citep{Alcock:1979} due to differential geometrical stretching along and transverse to the line of sight if the cosmological model used to translate redshifts to distances is incorrect. Measurement of these anisotropies therefore provides two correlated pieces of cosmological information. RSDs depend on the amplitude of the peculiar velocities and thus, in linear models, on the linear growth rate of structure $f(z)$, commonly parametrized in the combination $f(z)\sigma_8(z)$, while the AP effect constrains the combination $H(z)D_A(z)$, where $H(z)$ is the Hubble parameter and $D_A(z)$ the angular diameter distance at redshift $z$. The RSD and AP effects are in general degenerate. The Baryon Acoustic Oscillation (BAO) peak provides a clear, sharp feature in the galaxy clustering whose position is unaffected by RSD and this breaks the degeneracy and allows measurement of the AP effect. In other scenarios it is difficult to distinguish the AP effect without an accurate model for RSD (and vice versa). 

The theory of RSD in the galaxy clustering has been much studied \citep[e.g.][]{Scoccimarro:2004,Matsubara:2008,Taruya:2010,Reid:2011,Jennings:2011a} but is in general complicated by significant non-linear contributions even at quite large scales, and therefore requires sophisticated modelling. A recently proposed promising alternative has been to use RSD in the cross-correlation of galaxies with cosmic void centres in order to measure the growth rate \citep[e.g][]{Paz:2013,Hamaus:2015,Cai:2016a,Hawken:2017,Hamaus:2017a,Achitouv:2017a, Achitouv:2017b, Nadathur:2018a}. Voids have already proved useful cosmological tools in other scenarios, including for studying the integrated Sachs-Wolfe secondary anisotropies in the CMB \citep[e.g.][]{Granett:2008a,Hotchkiss:2015a,Nadathur:2016b,Kovacs:2016}, weak gravitational lensing \citep{Krause:2013,Melchior:2014,Clampitt:2015,Sanchez:2016} and the thermal Sunyaev-Zeldovich effect \citep{Alonso:2018}. A potential advantage of the use of voids for RSD studies is that galaxy dynamics in low density regions, where velocities are dominated by coherent bulk flows, can be modelled by linear theory alone. If the RSD effect around voids can be accurately modelled, they can be used in AP tests, as proposed by \citet{Lavaux:2012}, who argue that under certain circumstances voids may even outperform the BAO with future survey data. Some preliminary studies using voids in this way exist \citep{Sutter:2012tf,Mao:2017b} but current constraints are weak due to the difficulty of disentangling the AP and RSD effects.

A major advance in the modelling of the void-galaxy correlation in redshift space, $\xi^s(\mathbf{s})$, was made by \citet{Cai:2016a}, who provided a linear RSD model that matched simulation data well except in the deep interior regions of voids. \citet{Nadathur:2018a} extended this model by the addition of terms that are generally small for the \citet{Kaiser:1987} theory for the galaxy autocorrelation, but cannot be neglected within voids. The resulting completely linear model was shown to match simulation data extremely well on all scales, provided the real space positions of the voids could be determined. This opens the possibility of using voids to measure possible environment dependence of the growth rate, as well as to constrain cosmological parameters through a precise AP test.

However, one issue is that all current observational studies of RSD in the void-galaxy correlation \citep{Paz:2013,Hamaus:2015,Hawken:2017,Hamaus:2017a,Achitouv:2017a} make use of voids identified directly from redshift space galaxy data, in which case the real space void positions cannot be known. \citet{Chuang:2017} have recently pointed out that this raises difficulties for the theoretical modelling of $\xi^s(\mathbf{s})$, in general introducing an unknown velocity bias term on small scales. This is a consequence of a general theorem due to \citet{Seljak:2012}, and applies to RSD in other datasets as well, in particular the Lyman-$\alpha$ forest. 

In this work we elaborate on this problem by explicitly outlining several key common assumptions required for the derivation of \emph{all} current models for $\xi^s(\mathbf{s})$. These assumptions can be summarised as: the conservation of void numbers in redshift space, the invariance of void centre positions under the redshift space mapping, and the isotropy of the real space galaxy density and velocity fields around the void centres. These assumptions all necessarily hold if the positions of real space voids are known, but we demonstrate using simulation data that all of them fail for voids identified directly in the redshift space galaxy distribution. Our demonstration makes use of the watershed void-finding algorithm of \citet{Neyrinck:2008}, but we argue that these fundamental issues will also affect any other void finding routines. Thus changes to either the theoretical modelling or to the data analysis are required for a fully consistent analysis. 

We propose to achieve the latter: specifically, we use an algorithm to reconstruct the real space void positions from redshift space galaxy data based on the Zeldovich approximation \citep{Zeldovich:1970}, and now commonly used for BAO analyses \citep{Eisenstein:2007,Padmanabhan:2012}. The principle behind this method is to approximately reconstruct the real space galaxy distribution \emph{before} performing the void-finding step. We apply this method to mock galaxy and void catalogues based on an $N$-body simulation and demonstrate that it reproduces the agreement between theory and data previously seen for real-space selected voids by \citet{Nadathur:2018a}. We discuss the role of the assumed cosmological parameters in the reconstruction and show how the method can self-consistently be used to measure the growth rate $f$. The reconstruction also provides a method to measure the real space galaxy density profile of voids.

The layout of the paper is as follows. Section \ref{sec:data} describes the simulation data used, the creation of mock catalogues and the methods used to measure the correlation function. In Section \ref{sec:theory} we outline the derivation of the theoretical model for $\xi^s(\mathbf{s})$, discuss the assumptions made in the derivation of this and all other models, and show that these do not hold for voids in redshift space. Section \ref{sec:reconstruction} outlines the reconstruction technique we introduce to recover the real space void positions, and evaluates its performance on the simulation data. In Section \ref{sec:results} we provide an algorithm to self-consistently perform the reconstruction and measure the growth rate when $f$ is unknown, and show that this reproduces the fiducial value in our simulation. Appendix \ref{sec:appendix} provides a demonstration of the biases that can otherwise be introduced if the reconstruction procedure is not used. Finally we sum up and draw conclusions in Section \ref{sec:discussion}.

\section{Data}
\label{sec:data}

\subsection{Simulation and mocks}
\label{sec:mocks}
Simulation results in this paper make use of mock galaxy and void catalogues derived from the Big MultiDark (BigMD) $N$-body simulation \citep{Klypin:2016} from the MultiDark simulation project \citep{Prada:2012}. The BigMD simulation evolves $3840^3$ particles in a box of side $L=2500\;h^{-1}$Mpc using the {\small GADGET-2} \citep{Springel:2005} and Adaptive Refinement Tree \citep{Kravtsov:1997,Gottloeber:2008} codes. The cosmological parameters used for the simulation were $\Omega_M=0.307$, $\Omega_B=0.048$, $\Omega_\Lambda=0.693$, $n_\rmn{s}=0.95$, $\sigma_8=0.825$ and $h=67.8$, and initial conditions for the simulation were set using the Zeldovich approximation at starting redshift $z_i=100$. 

We use a halo catalogue created for the $z=0.52$ snapshot using the Bound Density Maximum algorithm \citep{Klypin:1997,Riebe:2013}, and populate these halos with mock galaxies using the Halo Occupation Distribution (HOD) model of \citet{Zheng:2007}, with parameters taken from \citet{Manera:2013} and designed to approximately reproduce the clustering and mean number density for galaxies in the Baryon Oscillation Spectroscopic Survey (BOSS) CMASS galaxy sample. Details of the algorithm, the assignment of central and satellite galaxies, and HOD model parameters used are described more fully in \citet{Nadathur:2017a}. This is the same mock galaxy catalogue as used by \citet{Nadathur:2018a}.

Dark matter (DM) densities in the simulation are measured using a $2350^3$ DM density grid generated from the full particle output of the simulation using a cloud-in-cell interpolation scheme. By comparing the power spectra of the clustering of DM and the mock galaxies at large scales, $k\lesssim0.05\;h\mathrm{Mpc}^{-1}$, we determine the linear bias value for the galaxy mocks, $b=1.87$. This is taken as the fiducial bias value in subsequent calculations.

We shift galaxy positions into redshift space, assuming the plane-parallel approximation and taking the line of sight direction to be along the $z$-axis of the simulation box, by applying the transformation 
\begin{equation}
\label{eq:s}
\mathbf{s} =\mathbf{x} + \frac{\mathbf{v}\cdot\hat{\mathbf{z}}}{aH}\,
\end{equation}
where $\mathbf{x}$ are the real space coordinates and $\mathbf{v}$ is the galaxy velocity.

Finally, using the redshift space galaxy field we also recover the `pseudo-real space' galaxy positions by using the reconstruction method described in detail in Section \ref{sec:reconstruction}. 

\begin{figure}
\begin{center}
\includegraphics[scale=0.4]{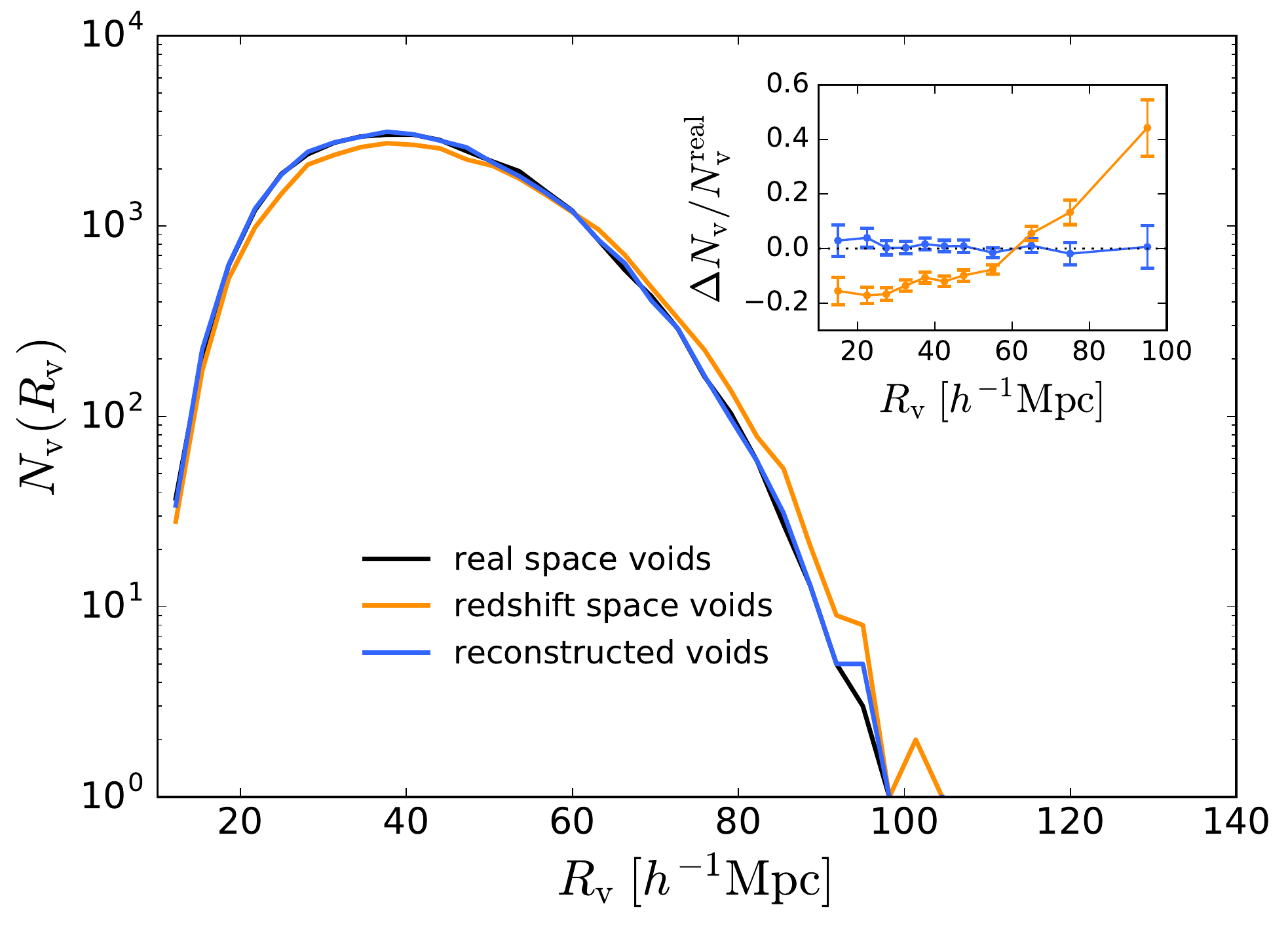}
\caption{The number of voids in the simulation as a function of void size $R_\mathrm{v}$, for voids found using galaxy positions in real space, redshift space, and reconstructed pseudo-real space. \emph{Inset}: The fractional difference in void numbers relative to real space voids, with the corresponding Poisson errors. The redshift space mapping leads to large changes at all void sizes. The reconstruction method recovers true void numbers to within $2\%$ in each $R_\mathrm{v}$ bin over the scales of interest (see text), and to better than $1\%$ overall.}
\label{fig:void_sizes}
\end{center}
\end{figure}

\subsection{Void catalogues}
\label{sec:voids}

\begin{table}
\centering
\caption{Total numbers of voids recovered in each galaxy population and differences relative to real space.}
\begin{tabular}{@{}lcc}
\hline
 & $N_\mathrm{v}$ & $(N_\mathrm{v}-N_\mathrm{v}^\mathrm{real})/N_\mathrm{v}^\mathrm{real}$ [\%] \\
\hline
Real space\ & 32742 & 0 \\
Redshift space\ & 29925 & $-8.6$ \\
Reconstructed space & 32913 & $+0.5$ \\
\hline\\
\end{tabular}
\label{table:voidnums}
\end{table}

We identify voids in the galaxy mocks described above through use of the \zobov watershed void-finding algorithm \citep{Neyrinck:2008}. The \zobov algorithm uses a Voronoi tessellation field estimator (VTFE) technique to reconstruct the galaxy density field from the discrete distribution, and then identifies local minima in this field and the watershed basins around them, to form a non-overlapping set of voids corresponding to local density depressions. We define each individual density basin as a distinct void, without any additional merging of neighbouring regions. A fuller description of the algorithm and void properties can be found in \citet{Nadathur:2016a} and \citet{Nadathur:2017a}.

We characterize each void by an effective spherical radius, $R_\mathrm{v}=\left(3V/4\upi\right)^{1/3}$, where $V$ is the total volume of the void -- note however that voids are in general of arbitrary shape and far from spherical. We assign the centre of each void to the centre of the largest sphere completely empty of galaxies that can be inscribed within the void \citep{Nadathur:2015b,Nadathur:2017a}.   
We apply the void-finding algorithm separately to the galaxy mocks in real space, redshift space and the reconstructed pseudo-real space. The size distributions of the resulting void populations are shown in Figure \ref{fig:void_sizes}, and the total void numbers are summarized in Table \ref{table:voidnums}.

The model for the RSD in the void-galaxy correlation is not expected to hold for all voids, in particular because galaxy velocities around small voids will be dominated by the local environment of structures outside the void rather than the void itself. We therefore restrict our attention to large voids. To achieve this, we apply the cut described in \citet{Nadathur:2018a} and remove all voids in each sample with effective radius $R_\mathrm{v}$ smaller than the median void size. Given the distributions shown in Figure \ref{fig:void_sizes}, this corresponds to selecting $R_\mathrm{v}\gtrsim43\;h^{-1}$Mpc, with slight variations for the different populations. All correlation function measurements presented in this work are calculated for void populations after the $R_\mathrm{v}$ cut has been applied. Note that as the cut is determined by the median void size in each sample, the fractional differences shown in Table \ref{table:voidnums} are unchanged by its application.

\subsection{Correlation function measurement and covariance matrix}
\label{sec:measurement}
To measure the void-galaxy correlation function we use the Landy-Szalay estimator \citep{Landy:1993} for cross-correlations,
\begin{equation}
\label{eq:landy-szalay}
\xi = \frac{D_1D_2-D_1R-D_2R+RR}{RR}\,
\end{equation}
where $D_1D_2$, $D_1R$ etc refer to the appropriately normalized number of pairs in a given separation bin. The random points $R$ are taken as a Poisson distributed set of points in the simulation box with $15\times$ the number density of the galaxies. 

Numerical implementation of Eq. \ref{eq:landy-szalay} is achieved using a modified version of the {\small CUTE} correlation function code \citep{Alonso:CUTE}\footnote{\url{http://members.ift.uam-csic.es/dmonge/CUTE.html}} to allow measurement of both the monopole $\xi(r)$ and $\xi(r,\mu)$, where $\mu$ is the cosine of the angle to the line of sight direction. We use 100 angular bins in the range $0\leq\mu\leq1$, and a radial bin width of $\Delta r=2.4\;h^{-1}$Mpc. From the measured values $\xi(r,\mu)$ we determine the appropriate quadrupoles
\begin{equation}
\label{eq:quad_defn}
\xi_2(r) = 5\,\int_0^1 \xi(r,\mu)P_2(\mu)d\mu\,,
\end{equation}
where $P_2(\mu)=\frac{1}{2}(3\mu^2-1)$ is the Legendre polynomial of order 2. The monopoles $\xi_0(r)$ are determined directly from Eq. \ref{eq:landy-szalay}.

The cross-correlation functions in Eq. \ref{eq:landy-szalay} are measured separately for each void catalogue with each galaxy population. In what follows, we will refer to the cross-correlations of voids with galaxies in real space, redshift space and reconstructed pseudo-real space using the notation $\xi^r$, $\xi^s$ and $\xi^p$, respectively. The particular void catalogue used in each case will be clear from the context.

As we use only a single simulation box, to estimate the error in our measurement of the multipoles we use the same jack-knife resampling technique as \citet{Nadathur:2018a}. We divide our simulation box into $N_{s}=512$ non-overlapping cubic sub-boxes, each measuring $312.5\;h^{-1}$Mpc on a side. We then determine the correlation function and multipoles excluding all the voids in each sub-box in turn, and combine these into the data vector $\mathbf{y}^{(k)}=(\xi^{(k)}_0,\xi^{(k)}_2)$, for $k=1,\ldots,N_s$. The covariance matrix is then determined as
\begin{equation}
\label{eq:covmat}
\mathbf{C}_{ij} = \frac{N_s -1}{N_s}\sum_{k=1}^{N_s}\left(y^{(k)}_i-\overline{y_i}\right)\left(y^{(k)}_j-\overline{y_j}\right)\,,
\end{equation}
where $\overline{y}$ denotes the mean of the jackknife samples. The $\chi^2$ value for a given theoretical model is then simply calculated as
\begin{equation}
\label{eq:chi2}
\chi^2 = \sum_{ij} \left(y^\mathrm{th}_i-y_i\right) \hat{\mathbf{C}}_{ij}^{-1} \left(y^\mathrm{th}_j-y_j\right)\,,
\end{equation}
where $\hat{\mathbf{C}}^{-1}$ is the unbiased estimator of the inverse covariance
\begin{equation}
\hat{\mathbf{C}}^{-1} = \frac{N_s-p-2}{N_s-1}\mathbf{C}^{-1}\,,
\end{equation}
including the Hartlap correction factor \citep{Hartlap:2007}, with $p$ being the length of the data vector.

\section{Theory}
\label{sec:theory}

\subsection{Linear RSD model for the void-galaxy correlation}
\label{sec:model}
A theoretical model for void-galaxy correlation in redshift space was derived by \citet{Nadathur:2018a}, extending that of \citet{Cai:2016a}, and providing a model that works on all scales including right into the void centres. In the following, we sketch out the main features of this model, while referring readers to \citet{Nadathur:2018a} for full details.

The derivation starts from the assumption that the number of void-galaxy pairs is conserved under the redshift-space mapping, so that \begin{equation}
\label{eq:conservation}
\left(1+\xi^s(\mathbf{s})\right)d^3s = \left(1+\xi^r(\mathbf{r})\right)d^3r\,,
\end{equation}
where $\mathbf{r}$ and $\mathbf{s}$ denote the vectors from the void centre to the galaxy position in real and redshift space respectively. If the void centre position does not suffer any RSD, it is only the galaxy velocity $\mathbf{v}$ that is relevant to the redshift space mapping from $\mathbf{r}$ to $\mathbf{s}$, 
\begin{equation}
\label{eq:coordinates}
\mathbf{s} = \mathbf{r} + \frac{\mathbf{v}\cdot\hat{\mathbf{X}}}{aH}\hat{\mathbf{X}}\,,
\end{equation}
where $a$ is the scale factor, $H$ the Hubble rate, and $\hat{\mathbf{X}}$ a unit vector in the line of sight direction to the void centre.

A further key assumption is that the average galaxy outflow velocity from the void is isotropic and radially directed, such that 
\begin{equation}
\label{eq:isotropy}
\mathbf{v} = v_r(r)\hat{\mathbf{r}}
\end{equation}
is a function of radial distance $r$ alone. For such a spherically symmetric case and where the density contrast of the void dominates over other structures in the environment, linear theory then predicts the coupling between the velocity field and the density to be 
\begin{equation}
\label{eq:vr}
v_r(r) = -\frac{1}{3}faH\Delta(r)r\,,
\end{equation}
where $\Delta(r)$ is the average mass density contrast within radius $r$ of the void centre,
\begin{equation}
\label{eq:Delta defn}
\Delta(r)\equiv\frac{3}{r^3}\int_0^r \delta(y)y^2 dy\,,
\end{equation}
with $\delta(r)$ the (isotropic) average mass density profile of the void, and $f=\mathrm{d}\ln D/\mathrm{d}\ln a$, with $D$ the growth factor and $a$ the scale factor, is the linear growth rate of density perturbations.

These equations can be combined to rewrite Eq. \ref{eq:conservation} as
\begin{equation}
\label{eq:xis basic}
1+\xi^s(\mathbf{s}) = \left(1+\xi^r(\mathbf{r})\right)\left[ 1+\frac{v_r}{raH} + \frac{\left(v_r^\prime-v_r/r\right)}{aH}\mu^2 \right]^{-1},
\end{equation}
where $\mu$ is the cosine of the  angle between the line-of-sight direction and the separation vector,
\begin{equation}
\label{eq:mu defn}
\mu \equiv \frac{\mathbf{X}\cdot\mathbf{r}}{|\mathbf{X}||\mathbf{r}|}=\cos\theta.
\end{equation}
This expression can be expanded to linear order in the densities $\delta$ and $\Delta$ to obtain the basic linear model for the redshift space correlation,
\begin{multline}
\label{eq:xis full}
\xi^s\left(s,\mu\right) = \xi^r(r)+\frac{f}{3}\Delta(r)\left(1+\xi^r(r)\right) \\ 
+f\mu^2\left[\delta(r)-\Delta(r)\right]\left(1+\xi^r(r)\right) \,,
\end{multline}
where the radial separations in real and redshift space are related by 
\begin{equation}
\label{eq:svr full}
r = s\left(1+\frac{f}{3}\Delta(s)\mu^2\right)\,.
\end{equation}
This is the equivalent of the Kaiser model \citep{Kaiser:1987} for RSD in the galaxy autocorrelation. Note that we do not assume a linear bias relationship within voids, i.e., $\xi^r(r)\neq b\delta(r)$ \citep[for a discussion of this point, see][]{Nadathur:2018a}\footnote{An alternative approach is taken by \citet{Pollina:2016}, who start by assuming such a linear bias relationship holds and then fit for $b$ taken as a free parameter. The recovered bias is strongly scale-dependent (see their Figure 4), meaning that a single bias value does not fit over all scales.}. 

Although this basic linear model captures much of the physics, to obtain an accurate fit to the data it is necessary to also account for the dispersion in galaxy velocities around the coherent outflow,
\begin{equation}
\label{eq:velocity disp}
\mathbf{v} = v_r\hat{\mathbf{r}} + v_{||}\hat{\mathbf{X}}\,,
\end{equation}
where $v_{||}$ is a zero-mean random variable with probability distribution function $P(v_{||})$. This results in an integral for the redshift space correlation,
\begin{equation}
\label{eq:xis dispersion}
1+\xi^{s,d}(\sigma,\pi) = \int dv_{||} P(v_{||})\left(1+\xi^s\left(\sigma,\pi-v_{||}/aH\right)\right),
\end{equation}
where $\sigma$ and $\pi$ are the redshift space distances transverse to and along the line of sight respectively, $r=\sqrt{r_\sigma^2+r_\pi^2}$, with $r_\sigma=\sigma$ and $r_\pi = \pi - (v_{||}+v_r\mu)/aH$, and the superscript $d$ differentiates the dispersion model from the one in Eq. \ref{eq:xis full}. We take the probability distribution function $P(v)$ to have a Gaussian form,
\begin{equation}
\label{eq:Pv}
P(v) = \frac{1}{\sqrt{2\upi}\sigma_v}\exp\left(-\frac{v^2}{2\sigma_v^2}\right)\,,
\end{equation}
with the dispersion a function of the real space radial separation scale, $\sigma_v=\sigma_v(r)$.

\citet{Nadathur:2018a} showed that for the case of voids selected in real space, for which the assumptions above are valid, Eq. \ref{eq:xis dispersion} provides an excellent fit to the measured redshift space void-galaxy correlation on all separation scales. We use this theoretical model for the main results presented in this paper. The fits with an earlier model by \citet{Cai:2016a} are considered in Appendix \ref{sec:appendix}.

\begin{figure*}
\begin{center}
\includegraphics[scale=0.52]{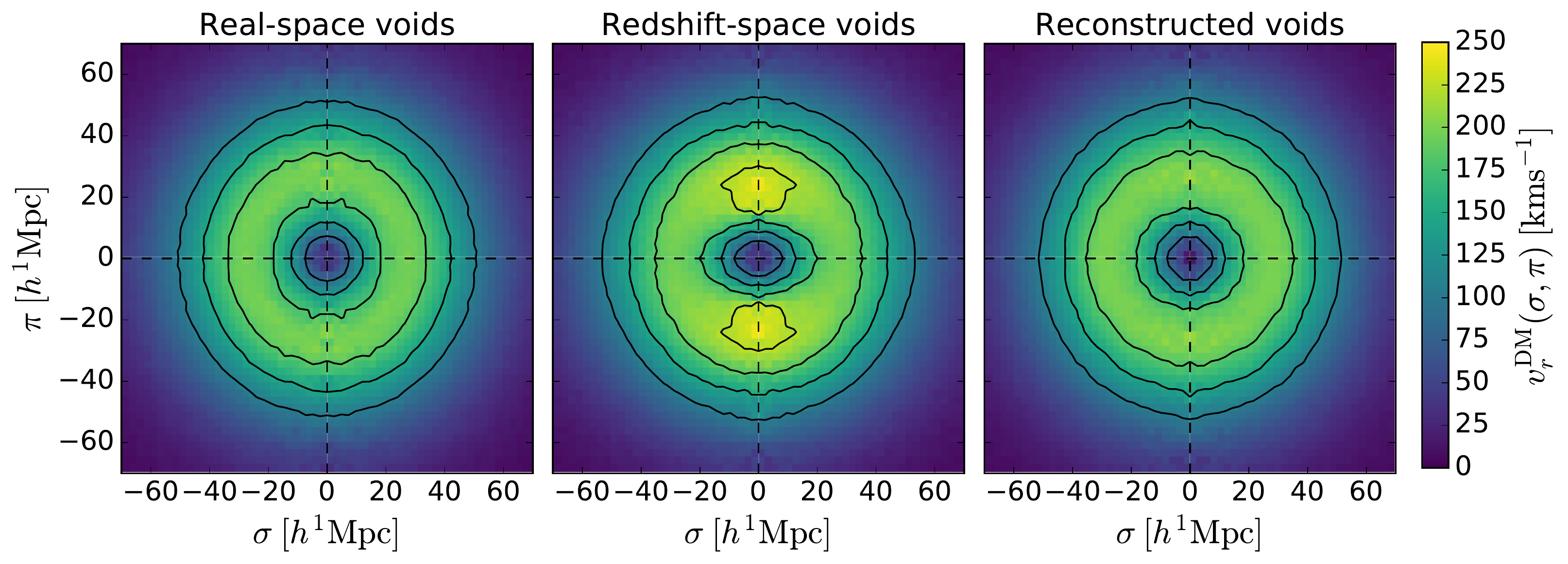}
\caption{The average radial component of the dark matter velocity outflow around void centers, $v_r^\mathrm{DM}(\sigma,\pi)$, as a function of real space distances from the void centre transverse to and parallel to the line-of-sight direction. Velocities are shown for voids identified using galaxy positions in real space (\emph{left}), redshift space (\emph{centre}) and reconstructed pseudo-real space (\emph{right}). The lines show contours of $v^\mathrm{DM}_r=100,\,140,\,180,\,220\;\mathrm{kms}^{-1}$ (the final contour only appears in the central panel). Contour levels were chosen to highlight differences between panels.}
\label{fig:2D velocity}
\end{center}
\end{figure*}

\subsection{Complications for voids based on redshift-space galaxies}
\label{sec:complications}
A number of theoretical models have now been proposed for RSD in the void-galaxy correlation function $\xi^s$ \citep[e.g.][]{Paz:2013,Cai:2016a,Achitouv:2017b,Nadathur:2018a}. They differ in derivation, and consequently they can lead to very different final predictions -- see \citet{Nadathur:2018a} for a discussion and comparison to simulation data.

Nevertheless, all models to date have been based on the same fundamental common assumptions highlighted in the previous section. These can be explicitly stated as: 
\begin{enumerate}
\item The number of voids is \emph{conserved} under the redshift space mapping (Eq. \ref{eq:conservation})
\item Void positions are \emph{invariant} under the redshift space mapping, so that the transformation $\mathbf{r}\rightarrow\mathbf{s}$ depends on galaxy velocities only (Eq. \ref{eq:coordinates})
\item The average radial outflow velocity around voids is \emph{isotropic} (Eq. \ref{eq:isotropy})
\item The real space correlation is \emph{isotropic}, $\xi^r(\mathbf{r})=\xi^r(r)$ 
\end{enumerate}
These conditions will necessarily hold if the real space void positions are known -- (i) and (ii) by construction, and (iii) and (iv) because there is no preferred direction along which any anisotropy could be introduced. 

However, all observational studies of the void-galaxy correlation until now have used voids identified directly in the redshift space galaxy distribution \citep{Paz:2013,Hamaus:2016,Hawken:2017,Hamaus:2017a,Achitouv:2017a}. In this scenario, each of these assumptions is violated, as we show below, leading to questions about the validity of the models.

The violation of void conservation in redshift space compared to real space, when using the same void-finding routine, has been previously noted \citep{Nadathur:2014a,Zhao:2016,Chuang:2017}. Comparison of the void populations in our simulation shows that the total number of voids changes by $\sim10\%$ under the redshift space mapping (Table \ref{table:voidnums}). Figure \ref{fig:void_sizes} shows that differences in void numbers apply over the entire range of void sizes $R_\mathrm{v}$ and are not limited to the smallest voids. In fact, the fractional difference relative to the number of real space voids is largest at large $R_\mathrm{v}$, where it can be as much as $40-50\%$. 

Assumption (ii) is equivalent to assuming that the void positions do not suffer RSD themselves. This has been shown to be incorrect for voids identified in redshift space galaxies by \citet{Chuang:2017}, who directly measure the quadrupole of the void autocorrelation function. In principle, it is possible to accommodate motion of the void centres by relaxing assumption (ii) and modifying Eq. \ref{eq:coordinates} to account for void velocities: this possibility is discussed by \citet{Cai:2016a}. However, as pointed out by \citet{Chuang:2017}, in general void-finding constitutes a non-linear transformation of the underlying galaxy (and thus, matter) density field, and therefore necessarily leads to a void velocity bias different from unity \citep{Seljak:2012}.\footnote{\citet{Chuang:2017} show that on very large scales, $s\gtrsim150\;h^{-1}$Mpc, the void density field $\delta_v$ reduces to a quasi-linear transformation of the galaxy density field $\delta_g$, and thus the void velocity bias on these scales is approximately unity. However, on the scales $s\lesssim100\;h^{-1}$Mpc relevant to the void-galaxy correlation this is not the case.} As the action of the void-finder algorithm cannot be described by a simple mathematical model, the exact form of the non-linear transformation is in general unknown and therefore so is the void velocity bias.  

Figure \ref{fig:2D velocity} shows the average radial component of the dark matter velocity field around void centres as a function of the transverse and line-of-sight distances from the centre, $v_r^\mathrm{DM}(\sigma,\pi)$, for different void populations.\footnote{We show profiles of $v_r^\mathrm{DM}$ rather than the average galaxy velocity as the latter are noisier close to the void centres due to low galaxy numbers. However, the strong anisotropy pattern for redshift space voids is present in both.} As expected, the velocity field is isotropic for real space voids. However, assumption (iii) clearly fails when voids are identified in the redshift space galaxy distribution. The physical reason for this is easy to understand intuitively: underdensities with higher line-of-sight outflow velocities will appear to have a lower central galaxy density in redshift space, and will therefore be preferentially selected by any void-finding algorithm, leading to the anisotropy in the stacked velocity profiles.

Note that the assumption of isotropy in the average velocity profile, $v_r=v_r(r)$, is separate from and more fundamental than the additional assumption of linear dynamics in Eq. \ref{eq:vr}. Indeed the latter assumption has been dropped in some works \citep{Achitouv:2017b} which nevertheless still assume isotropy.

\begin{figure}
\begin{center}
\includegraphics[scale=0.45]{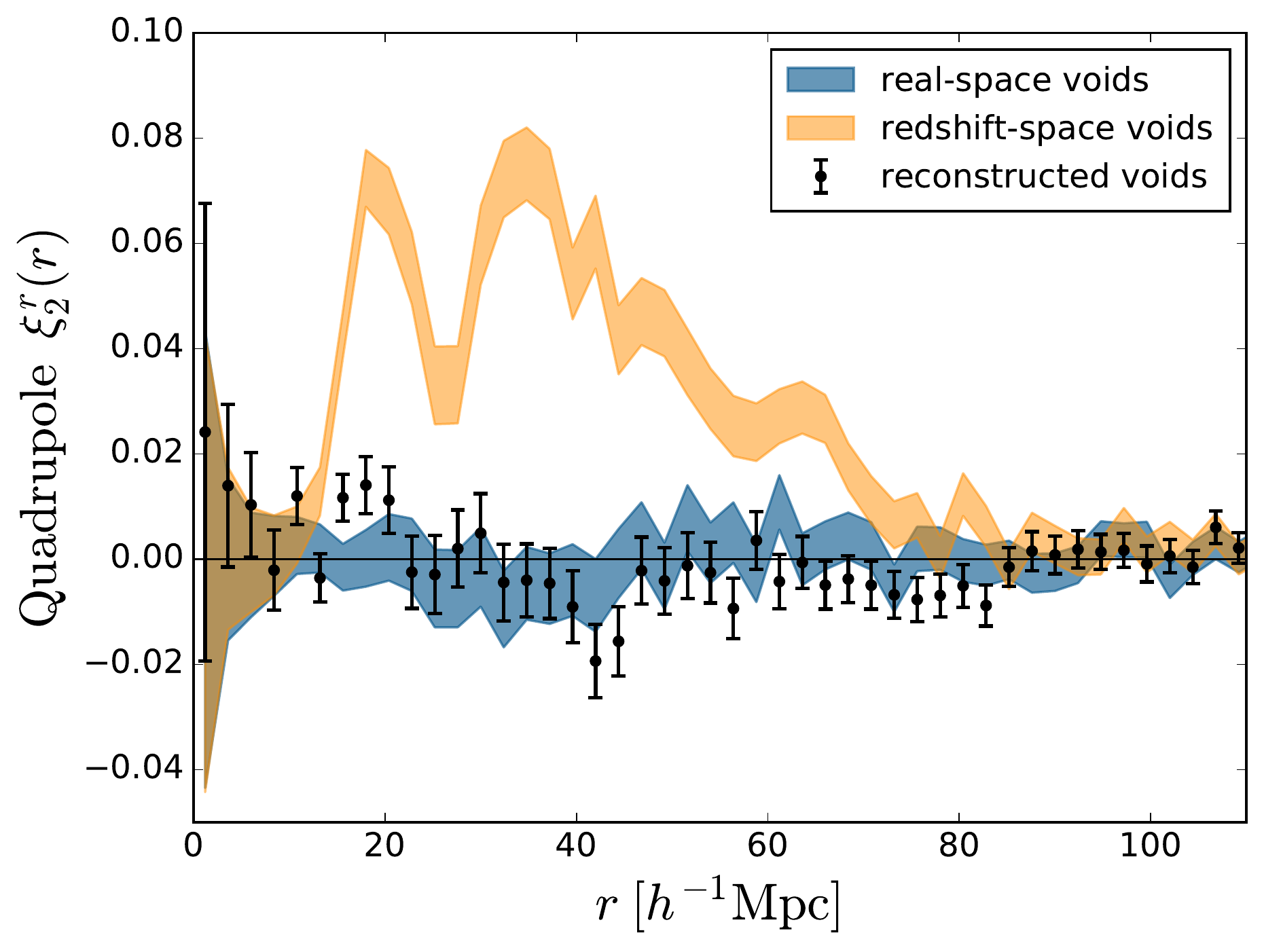}
\caption{Real space quadrupole $\xi^r_2(r)$ of the void-galaxy cross-correlation, for different void populations. The blue (yellow) shaded regions indicate the $1\sigma$ error range around the measured quadrupoles for real space (redshift space) voids. Data points with error bars indicate the measurements for voids found in the reconstructed pseudo-real space galaxy distribution. For real space and reconstructed voids $\xi^r_2(r)=0$ as required for the validity of the model, but this is not the case if voids are identified using redshift space galaxies.}
\label{fig:r-space_quad}
\end{center}
\end{figure}

The final assumption common to all models of RSD in the void-galaxy correlation is the isotropy of the correlation in real space, $\xi^r(\mathbf{r})$. To test this, we cross-correlate voids with the real space galaxy positions, and determine the quadrupole $\xi^r_2(r)$. The results are shown in Figure \ref{fig:r-space_quad}. For real space voids, $\xi^r_2(r)=0$ as required. However, when voids are identified using the redshift space galaxy positions, $\xi^r_2(r)$ is significantly non-zero over a range of scales. This is due to two contributing physical effects that are easily understood. Firstly, as assumption (ii) is violated and the redshift-space void positions have RSD themselves, their cross-correlation with a tracer without RSD (in this case the real space galaxy field) will still show an anisotropy. Secondly, void-finding in redshift space preferentially selects regions with larger outflow velocities along the line-of-sight direction as shown by Figure \ref{fig:2D velocity}. This exaggerates apparent underdensities along the line of sight; conversely in real space the contours of $\xi^r$ for such regions will be squashed along the line of sight, corresponding to $\xi^r_2>0$.

The results of this section show that all four of the fundamental assumptions required by \emph{all} models of the void-galaxy correlation including RSD proposed so far are violated for voids found by applying the void-finding algorithm directly on the redshift space galaxy distribution; thus none of the models currently existing in the literature are applicable to this scenario. The physical explanation for these results shows that the same problems will exist for any other void-finding algorithms applied directly to redshift space data.

Modifying the theoretical framework to account for redshift-space void selection appears a highly challenging task without a mathematical model of the action of the void-finder. On the other hand, \citet{Nadathur:2018a} showed that when the real space void positions are known and thus all the assumptions discussed are valid, the simple model of Eq. \ref{eq:xis dispersion} already provides an extremely good fit to simulation data. Therefore in the next section we describe an approach to reconstruct the real space void positions from redshift space galaxy data, rather than to alter the RSD model.

\section{Reconstruction method}
\label{sec:reconstruction}
In this section we describe the reconstruction method to recover the real space void positions from redshift space data. We first describe the physical basis of our method and the algorithm used, and then evaluate its effect on void-finding and the void-galaxy correlation.

\subsection{Understanding reconstruction}
\label{sec:recon_intro}
The goal of the reconstruction method described in the following is to recover the real space galaxy field using only the redshift space position information that would be available from survey data, and subtracting the effects of linear RSD from the galaxy positions. We will refer to the recovered galaxy positions as being in reconstructed or pseudo-real space. We then perform the void-finding step on the pseudo-real space galaxy field. To the extent that the RSD removal is successful, the voids thus obtained will match the true real space void positions, which are required for successful modelling of the anisotropic void-galaxy correlation. 

We then measure the cross-correlation of these reconstructed void positions with the original redshift space galaxy field to obtain $\xi^s(\mathbf{s})$. The cross-correlation of the void positions with the pseudo-real space galaxies gives $\xi^p(\mathbf{r})$, which we will show is a very good approximation of the true real space correlation $\xi^r(\mathbf{r})$. This provides a key input required for the theoretical modelling in redshift space, as well as being a quantity of interest in its own right \citep{Pisani:2014}.

The method used to solve for the galaxy displacement field is based on the concept of density field reconstruction \citep{Eisenstein:2007}, which is used frequently in the analysis of the BAO in the 2-point auto-correlation function of large redshift surveys \citep[e.g.][]{Padmanabhan:2012, Anderson:2012sa, Ross:2015, Alam:2017, Carter:2018}. The method makes use of first order Lagrangian perturbation theory to make an estimate of the linear displacement field from the observed evolved density field, and moves the over-density back along these vectors to approximately recover the linear over-density field \citep{Padmanabhan:2012}. 

Unlike in the case of the BAO reconstruction, we are only interested in the removal of the linear RSD contribution \citep{Kaiser:1987} and not the additional linearization of the evolved density field. For the BAO method the extra step is required to enhance the BAO peak which has been smoothed by non-linear density evolution and bulk flows, and leads to an improvement in precision of the BAO measurement of order $\sim 2$ at $z \sim 0$ \citep{Ross:2015}.

Other algorithms to recover the real space galaxy field from redshift space information have previously been studied \citep[e.g.][]{Wang:2009,Wang:2012,Kitaura:2012,Kitaura:2016a,Shi:2016}. Our method differs from these in its relative simplicity, as we only attempt to remove the effects of coherent large-scale flows and thus the linear RSD shifts. We show below that this is sufficient for our purposes, and leave to future work the investigation of improvements that may be obtained from more sophisticated approaches.

\subsection{Algorithm}
\label{sec:algorithm}

We work in a Lagrangian framework, in which the Eulerian position $\mathbf{x}$ at time $t$ is described in terms of the initial Lagrangian position $\mathbf{q}$ and a non-linear displacement vector $\mathbf{\Psi}(\mathbf{q}, t)$
\begin{equation}
\label{eq:Lagrangian framework}
\mathbf{x}(\mathbf{q}, t) = \mathbf{q} + \mathbf{\Psi}(\mathbf{q}, t)\,.
\end{equation}
To first order the overdensity field in Eulerian space, $\delta_{(1)}(\mathbf{x}, t)$, can be related to the displacement field in Lagrangian space, $\mathbf{\Psi}_{(1)}(\mathbf{q}, t)$, by
\begin{equation}
\label{eq:Zeldovich}
\mathbf{\nabla}_{\mathbf{q}}\cdot\mathbf{\Psi}_{(1)}(\mathbf{q}, t) = -\delta_{(1)}(\mathbf{x}, t).
\end{equation}
Assuming $\mathbf{\Psi}$ to be irrotational and applying a Fourier transform,
\begin{equation}
\label{eq:ZeldovichFT}
\mathbf{\Psi}_{(1)}(\mathbf{k}) = -\frac{i\mathbf{k}}{k^{2}}\delta_{(1)}(\mathbf{k}),
\end{equation}
which is the standard Zeldovich approximation \citep{Zeldovich:1970,White:2014b}.

In practice we observe the redshift space galaxy density field $\delta_g$. We assume that on large scales the matter and galaxy density fields are on average related by a linear bias $\delta_{g} = b\delta$. Eq. \ref{eq:Zeldovich} can then be modified to account for the RSD component \citep{Nusser:1994}
\begin{equation}
\label{eq:displacement}
\nabla\cdot\mathbf{\Psi}+\frac{f}{b}\nabla\cdot(\mathbf{\Psi}\cdot\mathbf{\hat{r}})\mathbf{\hat{r}} = -\frac{\delta_{g}}{b},
\end{equation}
where $f$ is the growth rate. Note that $1/b$ is required in the RSD term because dividing $\delta_{g}$ by $b$ on the RHS of this equation affects both the real-space and RSD components of the observed overdensity: this dependence was neglected in some previous papers, but included in others.

To compute the overdensity field we assign galaxies to a grid using a cloud-in-cell (CIC) algorithm. The resulting $\delta_g$ is then smoothed by convolving with a Gaussian filter in Fourier space, $S(k)=e^{{-(kR_s)^2}/2}$. The smoothing length $R_s$ must be chosen to be large enough to remove small-scale non-linearities but small enough to optimally capture information on the large-scale bulk flows. For BAO studies, typical smoothing lengths are $R_s=10-15\;h^{-1}$Mpc \citep{Padmanabhan:2012,Burden:2014,Achitouv:2015b,Vargas-Magana:2017}. Our fiducial results use a smoothing scale $R_s=10\;h^{-1}$Mpc. We discuss the optimal choice of $R_s$ further in Section \ref{sec:smoothing}.

We solve Eq. \ref{eq:displacement} using a fast Fourier transform (FFT) method \citep{Burden:2014, Burden:2015} for the full displacement field $\mathbf{\Psi}$. From this, we retrieve the RSD component \citep{Kaiser:1987,Padmanabhan:2012}
\begin{equation}
\label{eq:PsiRSD}
\mathbf{\Psi}_{\mathrm{RSD}} = -f(\mathbf{\Psi}\cdot\mathbf{\hat{r}})\mathbf{\hat{r}},
\end{equation}
and shift the individual galaxy positions by $-\mathbf{\Psi}_\mathrm{RSD}$ to approximately remove the linear RSD on each galaxy. The computational expense of the FFT-based reconstruction method is comparable to that of the void-finding step.

In order to apply reconstruction we require input values for the galaxy bias $b$ and the growth rate $f$. For the fiducial model, these values are $b=1.87$ and $f=0.761$, determined appropriately for our galaxy mocks and the simulation redshift. However, part of the motivation for measuring the void-galaxy correlation in redshift space is precisely to determine the true value of $f$. In Section \ref{sec:recon_cosmology} we therefore consider the effects on $\xi^s$ of performing the reconstruction with the wrong assumed cosmology, and in Section \ref{sec:results} we outline a non-circular, self-consistent method for the measurement of $f$ where we allow the model to be tested to dictate what value of $f$ we should assume in the reconstruction step, and show that when applied to our simulation data it correctly returns the fiducial growth rate after marginalization over $b$.

It is worth commenting on the fact that for the reconstruction we assume the validity of a constant linear galaxy bias $b$. Within voids it has been argued that such a bias relationship does not hold (\citealt{Neyrinck:2014,Nadathur:2015b,Nadathur:2015c,Nadathur:2018a}) and we do not assume a linear bias for the determination of $\delta(r)$ in Section \ref{sec:model}. This failure of the linear bias within voids can be explained purely as a statistical selection effect in the presence of stochasticity in the bias relationship, due to the restriction to regions specifically chosen to have very few galaxy tracers, and thus low $\delta_g$. On the other hand, when considering all values of $\delta_g$ over the entire simulation box, \emph{on average} the linear bias relationship $\langle\delta|\delta_g\rangle=\delta_g/b$ remains a good approximation. The use of the CIC grid assignment and the Gaussian smoothing in reconstruction also help to suppress statistical deviations from the mean relationship.

\begin{figure*}
\begin{center}
\includegraphics[scale=0.45]{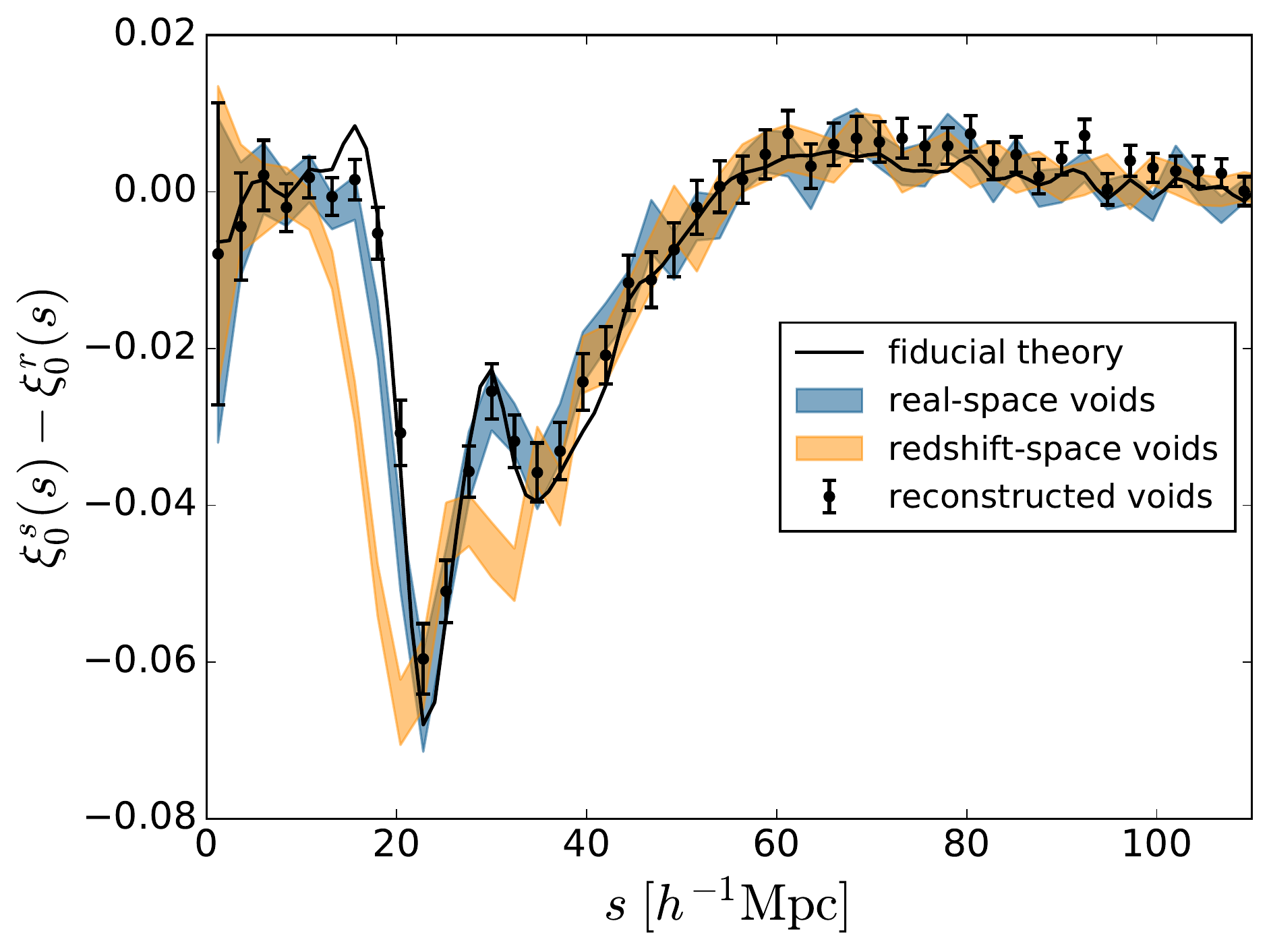}
\includegraphics[scale=0.45]{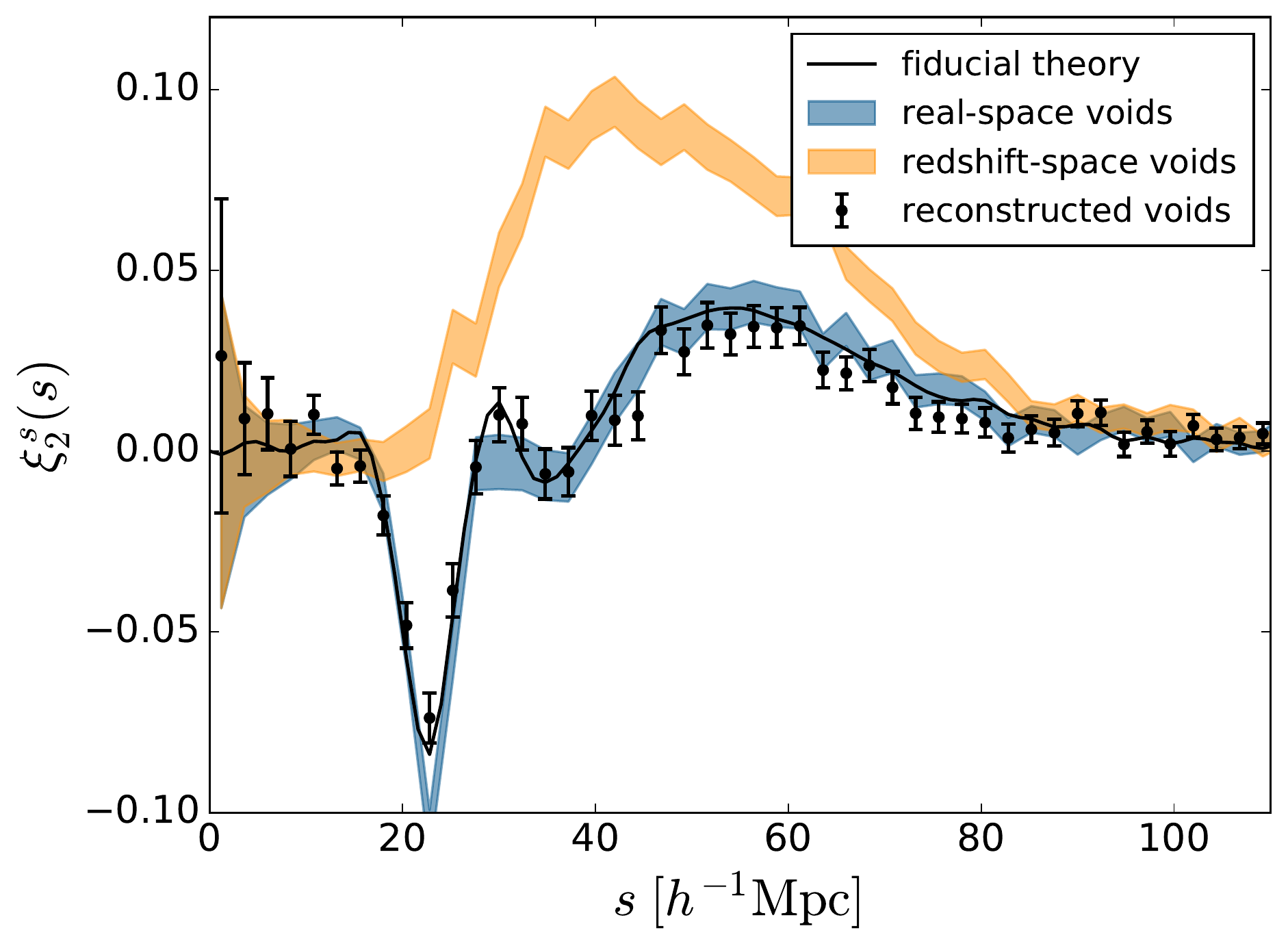}
\caption{Measured multipoles for the void-galaxy correlation with galaxies in redshift space (data points with error bars), compared to the fiducial theory prediction from Eq. \ref{eq:xis dispersion} (solid line). Voids were identified from the redshift-space galaxy data using the reconstruction method described in Section \ref{sec:reconstruction}. \emph{Left}: monopole, shown as difference to the real space monopole for clarity. \emph{Right}: Quadrupole. Shaded regions in both panels show the corresponding measured multipoles and $1\sigma$ error range for voids identified in redshift space without reconstruction.}
\label{fig:fiducial_recon}
\end{center}
\end{figure*}

\subsection{Effect of reconstruction on void-finding}
\label{sec:recon_void_finding}
Table \ref{table:voidnums} shows that the total number of voids identified in the reconstructed pseudo-real space galaxy field agrees with the number of real space voids to within $0.5\%$ when the reconstruction is performed in the fiducial cosmology. The size distribution of these voids also closely matches that of real space voids, as shown in Figure \ref{fig:void_sizes}. The fractional difference in void numbers, $\Delta N_\mathrm{v}/N_\mathrm{v}^\mathrm{real}$, is almost constant as a function of void size $R_\mathrm{v}$, with the largest differences arising at the smallest void sizes. For the larger voids we consider for correlation function measurements, $R_\mathrm{v}\gtrsim43\;h^{-1}$Mpc, the difference in numbers is $<2\%$ in each individual $R_\mathrm{v}$ bin, and $0.5\%$ overall. We therefore conclude that the reconstruction of redshift space data does conserve void numbers to a very good approximation, and thus that assumption (i) of Section \ref{sec:complications} is valid.

The third panel in Figure \ref{fig:2D velocity} shows the radial velocity outflow patterns around reconstructed void positions. In contrast to the case of redshift space voids, these profiles are isotropic and very similar to those for the true real space voids. Thus reconstruction is able to recover the validity of the assumption (iii), that $v_r(\mathbf{r})=v_r(r)$.

Finally, we test the isotropy of the cross-correlation of reconstructed void positions with the real space galaxy distribution, $\xi^r(\mathrm{r})$. Figure \ref{fig:r-space_quad} shows the quadrupole $\xi^r_2(r)$ for this correlation (the data points with error bars), which is consistent with zero to good approximation. This verifies assumption (iv) made in Section \ref{sec:complications}.

The validity of assumption (ii) remains to be tested. In the following we will do this indirectly, by comparing the measured redshift space correlation of reconstructed voids to the theory derived from using this assumption. Note however that if residual RSD effects were present in the reconstructed void positions, these would also contribute to $\xi^r_2(r)\neq0$. Thus Figure \ref{fig:r-space_quad} already gives good reason to be confident that assumption (ii) is also valid.

\subsection{Effect of reconstruction on $\xi^s$}
\label{sec:recon_xi}
We now compare the model of Section \ref{sec:model} to the measured redshift space correlation function multipoles $\xi^s_0$ and $\xi^s_2$ for reconstructed voids. 

To calculate the theory model using Eqs. \ref{eq:xis dispersion} and \ref{eq:xis full} requires specifying three functions in real space: $\xi^r(r)$, $\delta(r)$ and $\sigma_v(r)$. The correlation with real space galaxies is not known; however, the reconstruction provides a practical alternative, which is to directly measure the correlation of voids with the pseudo-real space galaxy field, $\xi^p_0(r)$. We tested the differences between $\xi^r_0$ and $\xi^p_0$ in the simulation data, and found that they were small, meaning that the method allows accurate reconstruction of the true real-space void galaxy density profile \citep{Pisani:2014}. We use the measured $\xi^p_0(r)$ as the input for the theory predictions.

The functions $\delta(r)$ and $\sigma_v(r)$ cannot be directly determined from galaxy survey data, but must be calibrated from mock void and galaxy catalogues in simulations. We measure the stacked average profiles of $\delta(r)$ and $\sigma_v(r)$ for the real and reconstructed void populations and find only small differences, which when propagated through the theory lead to negligible differences in the predictions for $\xi^s_0(s)$ and $\xi^s_2(s)$. We therefore fix the functions to match the measured values for real space voids, as done by \citet{Nadathur:2018a}.

Figure \ref{fig:fiducial_recon} shows the predictions for $\xi^s_0$ and $\xi^s_2$ for the fiducial growth rate $f=0.761$, together with the measured multipoles for cross-correlation of each of the real space, redshift space and reconstructed void populations with redshift space galaxies. For visual clarity, the monopoles are shown as the difference with respect to the real space monopole $\xi^r_0(r)$. Physical explanations for the various features seen -- in particular the negative dip and positive peak in the quadrupole -- were discussed by \citet{Nadathur:2018a}. The correlation functions for real space and reconstructed voids agree very well with each other: this is evidence that the reconstruction is successfully recovering the real space void positions. The model of Eq. \ref{eq:xis dispersion} also provides an excellent description of the data: the reduced $\chi^2$ value for fits to $\xi^s_0(s)$ and $\xi^s_2(s)$ for reconstructed voids is 1.01 for the fiducial $f$. 

Figure \ref{fig:fiducial_recon} also shows that the void-galaxy correlation is very different for redshift-space voids, with particularly striking differences for the quadrupole. $\xi^s_0$ and $\xi^s_2$ for redshift space voids match neither the data for real space voids, nor the theory predictions. In Appendix \ref{sec:appendix} we consider a different theory model and show that it also fails to describe the data for redshift space voids. This is unsurprising given the discussion in Section \ref{sec:complications}: the use of redshift space voids invalidates the key assumptions made in deriving \emph{all} theoretical models for $\xi^s$.

\begin{figure}
\begin{center}
\includegraphics[scale=0.42]{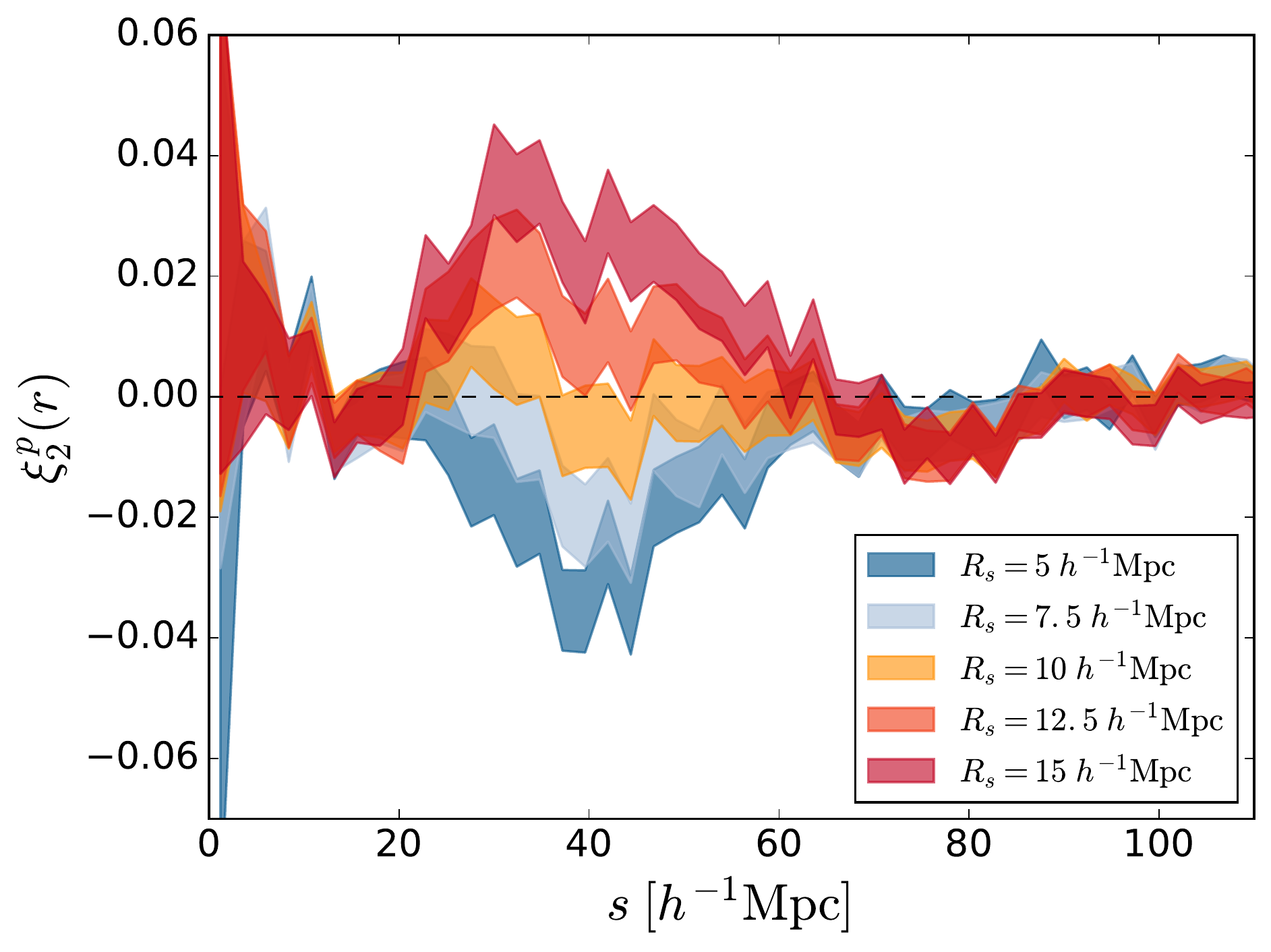}
\caption{The quadrupole $\xi^p_2(r)$ for the correlation of reconstructed void and galaxy positions, for different choices of the fixed smoothing scale $R_s$ used in the reconstruction procedure (see text). The choice $R_s=10\;h^{-1}$Mpc results in an isotropic correlation function consistent with expectation $\xi^{p,\mathrm{th}}_2(r)=0$, but residual anisotropies are seen for other values.}
\label{fig:smoothing}
\end{center}
\end{figure}

\subsection{Choice of smoothing scale}
\label{sec:smoothing}
Our implementation of the reconstruction algorithm requires a choice of width $R_s$ for the Gaussian filter used to smooth the galaxy density field. Reconstruction implementations for BAO measurements typically use a fixed smoothing scale $R_s\simeq10-15\;h^{-1}$Mpc \citep{Padmanabhan:2012,Burden:2014,Achitouv:2015b}. $R_s$ must be large enough to ensure the validity of the Zeldovich approximation used for Eq. \ref{eq:displacement}, but the appropriate value is not known \emph{a priori}. If $R_s$ is too large, some of the contributions to Eq. \ref{eq:displacement} from density fluctuations on linear scales are missed, so the reconstruction will fail to remove all of the linear RSD from the galaxy distribution and the reconstructed voids will not match the true real space voids. If $R_s$ is too small, the reconstruction picks up contributions from small scale density fluctuations for which Eq. \ref{eq:displacement} is not accurate; again the reconstruction of void positions will fail to satisfy the conditions discussed in Section \ref{sec:complications}.

To explore this choice, we empirically tested the performance of 5 different values of the smoothing scale in the range $5\leq R_s\leq15\;h^{-1}$Mpc. For each choice we re-ran the reconstruction and the void-finding, and measured the quadrupole of the cross-correlation of these voids with the reconstructed galaxy field, $\xi^p_2(r)$. If the reconstruction procedure successfully removes all anisotropic effects and recovers the real space void positions, this quadrupole should be $\xi^{p,\mathrm{th}}_2(r)=0$. On the other hand, failure to reproduce any of the key conditions of Section \ref{sec:complications} would be expected to introduce a non-zero quadrupole.

Figure \ref{fig:smoothing} shows the results for different $R_s$. The fiducial choice of smoothing scale, $R_s=10\;h^{-1}$Mpc, results in a quadrupole consistent with zero, as already discussed. As expected, larger values of $R_s$ indeed fail to remove all of the anisotropy, giving results intermediate between those of real space and redshift space voids seen in Figure \ref{fig:r-space_quad}. For $R_s<10\;h^{-1}$Mpc, the reconstruction over-corrects in the opposite direction, leading to $\xi^p_2<0$ at scales $r\sim40\;h^{-1}$Mpc. Fits to these data of the expected isotropic model $\xi^{p,\mathrm{th}}_2(r)=0$ give reduced $\chi^2$ values of 2.0, 1.3, 0.8, 1.3 and 2.3 for $R_s=5,\;7.5,\;10,\,12.5$ and $15\;h^{-1}$Mpc respectively. Based on these results, we set $R_s=10\;h^{-1}$Mpc for all the other reconstruction results reported in this paper.

It should be noted that the need to choose $R_s$ in this manner is a limitation of current reconstruction algorithms that use a fixed smoothing scale. Iterative determination of the appropriate smoothing scale \citep[e.g.][]{Schmittfull:2017,Hada:2018} provides a promising alternative which might lead to further improvements, but we defer such investigations to future work.

\begin{figure*}
\begin{center}
\includegraphics[scale=0.55]{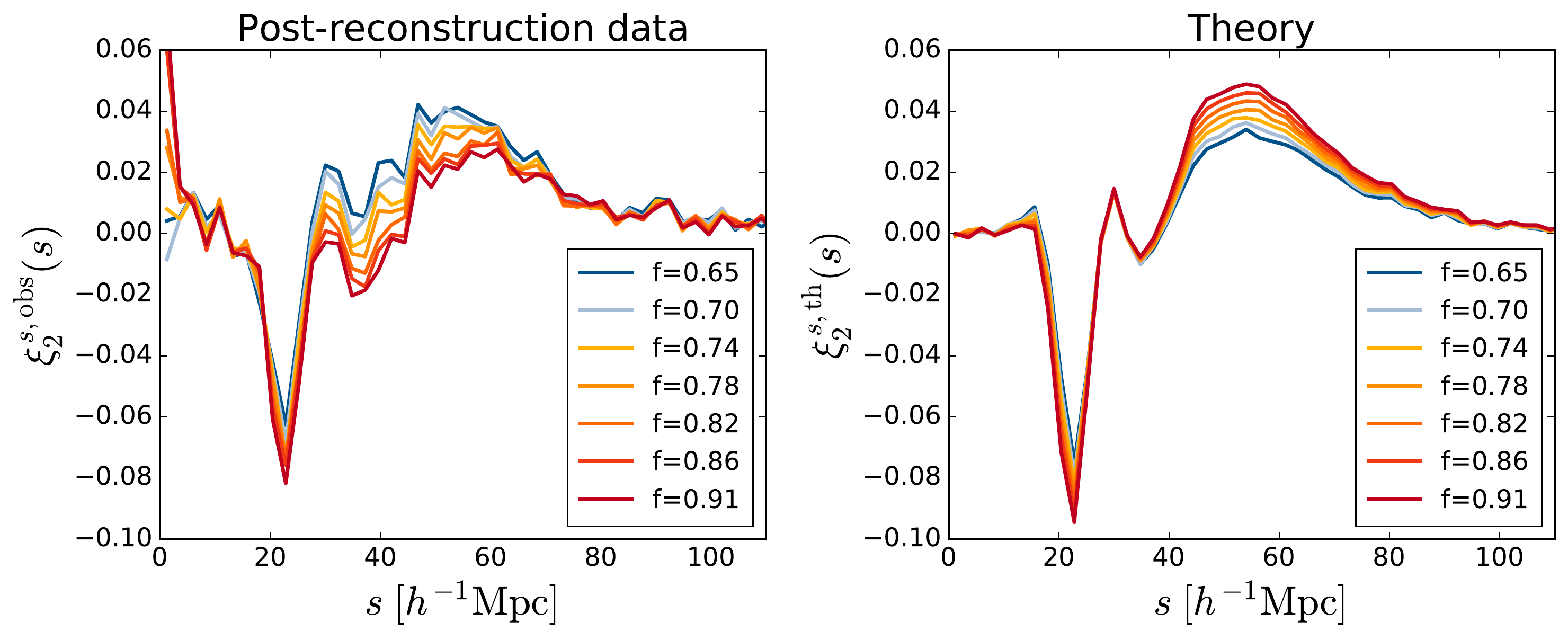}
\caption{\emph{Left}: Effect on the measured redshift space quadrupole $\xi^s_2(s)$ when the assumed growth rate $f$ is changed in the reconstruction procedure prior to void-finding. Error bars are omitted for clarity. \emph{Right}: Effect on the theoretical predicted quadrupole from changes in $f$. For these plots the assumed bias is kept fixed at the fiducial value, $b=1.87$.}
\label{fig:varying_recon_quad}
\end{center}
\end{figure*}

\begin{figure}
\begin{center}
\includegraphics[scale=0.45]{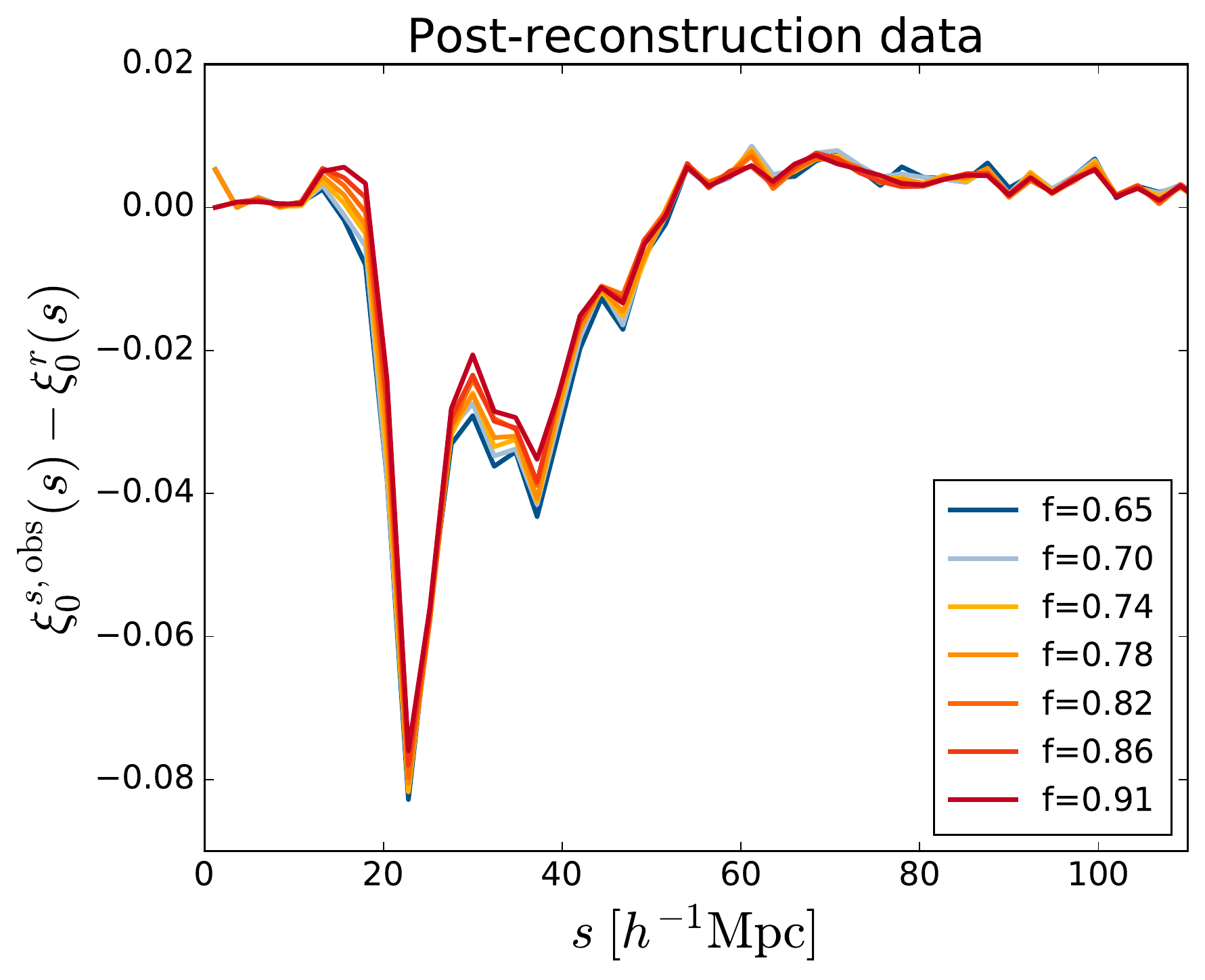}
\caption{Effect on the measured redshift space monopole $\xi^s_0(s)$ when the assumed growth rate $f$ is changed in the reconstruction procedure prior to void-finding. For visual purposes, the data is shown as the difference relative to a fixed fiducial function $\xi^r_0(s)$. Error bars are omitted for clarity. The assumed bias is kept fixed, $b=1.87$.}
\label{fig:varying_recon_mono}
\end{center}
\end{figure}

\subsection{Dependence of reconstruction on assumed cosmology}
\label{sec:recon_cosmology}
In addition to $R_s$, implementation of the reconstruction algorithm requires the specification of parameters $f$ and $b$, the growth rate and linear galaxy bias. The growth rate is cosmology-dependent and is not known \emph{a priori} for survey data; indeed an important motivation for measuring the void-galaxy correlation is to determine $f$. We therefore consider the effects on $\xi^s$ of performing the reconstruction with the wrong input growth rate. 

For each input value of $f$, we re-perform the reconstruction of the pseudo-real space galaxy field and the subsequent void-finding, and then measure the cross-correlation of these voids with the redshift space galaxies, $\xi^s$. The left panel of Figure \ref{fig:varying_recon_quad} shows the variation of the resultant quadrupole, $\xi^{s,\mathrm{obs}}_2$, for selected values of $f$. A significant trend can be seen in the height of the peak in $\xi^{s,\mathrm{obs}}_2$ in the range $30\lesssim s\lesssim 60\;h^{-1}$Mpc, as well as a smaller trend in the depth of the dip at $s\sim25\;h^{-1}$Mpc. This trend can be heuristically understood as follows. When the assumed $f$ is smaller than the true growth rate, the reconstruction fails to remove all of the RSD in the void positions. Thus $\xi^{s,\mathrm{obs}}_2$ is closer to that for the redshift space voids shown in Figure \ref{fig:fiducial_recon} (which corresponds to the limit of assuming $f=0$ in the reconstruction). In particular this means a smaller dip and a larger peak in the quadrupole. Conversely, if the assumed $f>f_\mathrm{true}$, void positions are over-corrected and an opposite RSD effect is introduced in $\xi^{s,\mathrm{obs}}_2$, increasing the depth of the dip and decreasing the height of the peak.

The right panel of Figure \ref{fig:varying_recon_quad} shows the corresponding changes to the predicted quadrupole $\xi^{s,\mathrm{th}}_2$ as $f$ is changed. Note that each prediction accounts for not only the direct effect of changes in $f$, but also the smaller effect of changes in the monopole $\xi^p_0(s)$, which is determined separately for each reconstruction. Importantly, the trend of changes in the peak height around $s\sim50\;h^{-1}$Mpc is opposite to that in $\xi^{s,\mathrm{obs}}_2$, i.e., larger assumed $f$ leads to a higher predicted peak but a lower observed one. Only for values of the assumed growth rate close to $f_\mathrm{true}$ do the theory and data match.

Figure \ref{fig:varying_recon_mono} shows the corresponding effect on the observed monopole, $\xi^{s,\mathrm{obs}}_0(s)$. To visually highlight the differences, the same fiducial function $\xi^r_0(s)$ has been subtracted from each observed monopole. Changes to the $f$ assumed in reconstruction produce smaller changes in $\xi^{s,\mathrm{obs}}_0(s)$ than in the quadrupole, and over a more restricted range of scales $s$. For the predicted monopole, the effects of changing the measured input monopole $\xi^p_0$ with each reconstruction are larger than for the quadrupole. Interestingly, they also act in the opposite direction to the effect of changing $f$. As a result, the predictions for $\xi^{s,\mathrm{th}}_0(s)$ have a very small dependence on the assumed $f$, and so are not shown.

\section{RSD fitting and results}
\label{sec:results}

\begin{figure}
\begin{center}
\includegraphics[scale=0.55]{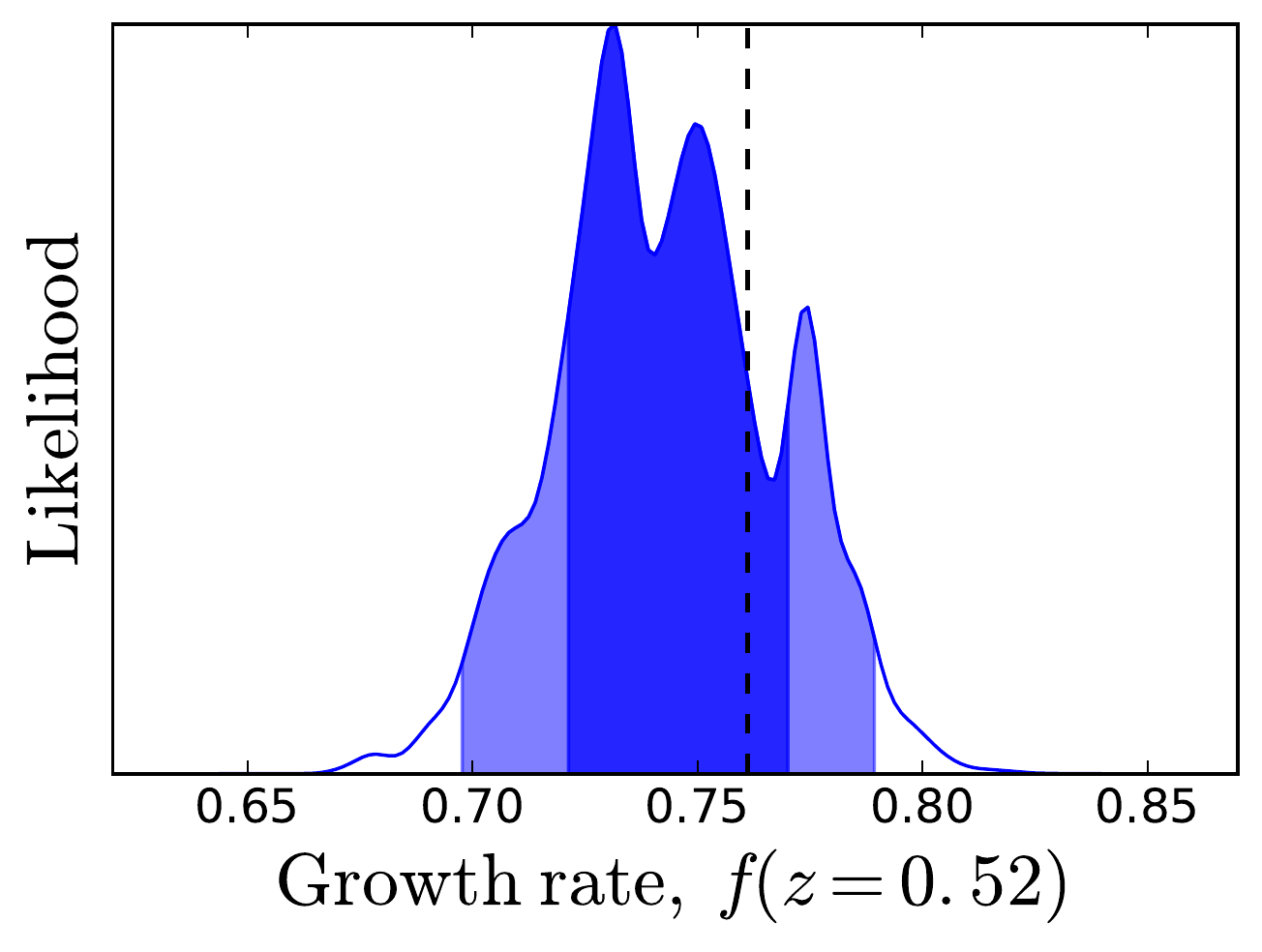}
\caption{The likelihood for the growth rate $f$, after marginalization over $b$, obtained from fits to the redshift space multipoles $\xi^s_0(s)$ and $\xi^s_2(s)$ of the void-galaxy correlation, using the reconstruction procedure described in the text. The recovered mean value is $f=0.744^{+0.027}_{-0.024}$ (68\% c.l.), consistent with the fiducial value $f_\mathrm{fid}=0.761$ (vertical dashed line).}
\label{fig:likelihood}
\end{center}
\end{figure}

The results of Section \ref{sec:recon_cosmology} suggest that the reconstruction method can be used to measure the true growth rate $f$ from measurement of the void-galaxy correlation $\xi^s$. To ensure that this process is performed self-consistently, we adopt the following algorithm:
\begin{enumerate}
\item To perform the reconstruction, assume a set of cosmological parameters; for our plane-parallel simulation setup, this requires a choice of the pair $(f,b)$;
\item For each such choice, reconstruct the pseudo-real space galaxy field, run the void-finding step, and apply the void size cut to obtain the sample for cross-correlation;
\item Cross-correlate these void positions with the reconstructed galaxy positions to measure $\xi^p_0(r)$, and with the redshift space galaxy positions to measure $\xi^s_0(s)$ and $\xi^s_2(s)$;
\item Using the measured $\xi^p_0(r)$ and the assumed $f$, obtain the predicted $\xi^{s,\mathrm{th}}(\mathbf{s})$ using the dispersion model of Eq. \ref{eq:xis dispersion}, and decompose it into multipoles;
\item Obtain the $\chi^2$ for the fit of this model to the data (Eq. \ref{eq:chi2});
\item Return to step (i) and repeat for a different choice of $(f,b)$. 
\end{enumerate}

For the purposes of this work, $(f,b)$ are the only relevant parameters and so a simple grid search over parameter space is sufficient. We assume flat prior ranges of $f\in(0.61,0.91)$ and $b\in(1.77,1.95)$. For simplicity, we keep $\delta(r)$ and $\sigma_v(r)$ fixed to the values determined for the real space voids: empirically we found that these change negligibly for reconstructed voids and do not depend on the choice of $(f,b)$ in the reconstruction. We also keep the covariance matrix $\mathbf{C}_{ij}$ fixed, under the assumption that it does not depend strongly on reconstruction parameters.

For application of this technique to actual survey data, other cosmological parameters will need to be included, in particular $\Omega_m$, $\sigma_8$ and parameters $\alpha_\perp$ and $\alpha_{||}$ to account for Alcock-Paczynski distortions. This will also require testing the dependence of $\delta(r)$ and $\sigma_v(r)$ on $\Omega_m$ and $\sigma_8$, which is beyond the scope of this work.

The method outlined here results in changes to both theory model \emph{and} data as the parameters are changed. This is necessary to self-consistently fit for the growth rate; however, an unfortunate and unavoidable consequence is that the resultant likelihood surface is noisy. Figure \ref{fig:likelihood} shows the likelihood for the recovered growth rate $f$ after marginalization over the bias, corresponding to a mean recovered value $f=0.744^{+0.027}_{-0.024}$ ($68\%$ c.l.). This is consistent with the fiducial value $f_\mathrm{fid}(z=0.52)=0.761$ for the simulation, and corresponds to a $\sim3.4\%$ measurement of the growth rate. The noise introduced by the reconstruction is reflected in the fact that the likelihood does not behave as a smooth function for small changes in $f$, and the best-fit value is slightly lower, $f_\mathrm{best}=0.732$. This can be contrasted with the smoother likelihood surface obtained by \citet{Nadathur:2018a} when the real space void positions were assumed to be known exactly and reconstruction was not required.

\section{Discussion}
\label{sec:discussion}

Redshift space distortions to the void-galaxy correlation are extremely interesting for three reasons: they can provide a measure of the linear growth rate complementary to that from galaxy clustering, that is sensitive to environmental variations in $f$ in low density regions as is expected in some modified gravity scenarios; they can be accurately modelled on all separation scales -- at least when the real space void positions can be determined -- using linear theory alone; and the success of linear models for the RSD in $\xi^s$ potentially allows the use of voids in Alcock-Paczynski tests of cosmology that probe a different and wider range of scales than those using the BAO peak. Thus, RSD measurements using voids will therefore be important tools in the analysis of large-scale structure data from future galaxy redshift surveys.

However, while linear models of $\xi^s$ describe simulation data well, their applicability to the case where voids are identified directly from redshift space galaxy positions was questioned by \citet{Chuang:2017}. We addressed this from first principles, identifying four key assumptions common to all existing theoretical models of $\xi^s$ and explicitly showing that each of them is violated if void-finding is performed on the redshift space galaxy field. Modifying the theory to account for this appears to be a hard problem. Instead we have proposed a practical approach: applying the reconstruction technique to obtain real space voids from redshift space data, based on using the Zeldovich approximation to approximately recover the real space galaxy distribution. If this step is performed prior to void-finding, the validity of the model assumptions is restored and the linear theory model \citep{Nadathur:2018a} provides an excellent fit to simulation data on all scales. 

Performing the reconstruction requires input values for the cosmological parameters, in particular the growth rate $f$, which we wish to measure. In order to make our analysis non-circular, we need to re-perform reconstruction for every model tested, assuming that model represents the true cosmology. We find that different choices of $f$ affect the reconstructed data and theory in different ways, so a simple self-consistent algorithm can be used to measure $f$ from the reconstructed data. The technique also allows us to easily measure the real space void galaxy density profile. Applying this method to mock data from the Big MultiDark simulation, we were able to recover a $3.4\%$ measurement of the growth rate at $z\sim0.5$ after marginalization over the bias, in good agreement with the fiducial value. The reconstruction process introduces noise, however, and so the recovered likelihood is not smooth for small changes in the model.

The use of \emph{some} method to recover real space voids from redshift space data is necessary for the consistent application of any theoretical models to measurements of the void-galaxy correlation. The reconstruction method outlined here works sufficiently well, but further improvements are still possible. The choice of a fixed smoothing scale $R_s$ is not ideal: the implementation of an iterative determination of the best smoothing scale \citep{Schmittfull:2017,Hada:2018} would be better. It is also possible that other methods of estimating the velocity field from galaxy data, particularly Bayesian methods \citep{Jasche:2013,Leclercq:2015b,Kitaura:2016a}, could improve the correspondence between the reconstructed and true real space voids, leading to better agreement with theory. 

The additional computational expense of performing the reconstruction is comparable to that of the tessellation required for the void-finding. For current data volumes, these steps can be run on a standard PC in a few minutes. Repeated iteration in order to derive the likelihood as described in Section \ref{sec:results} is feasible using parallel computing on a cluster. Unfortunately, as discussed in Appendix \ref{sec:appendix}, this additional expense is necessary for a self-consistent analysis.

A generalization of the method presented here to apply to survey data requires the introduction of several more free parameters than were considered for the simulation case, in particular the Alcock-Paczynski parameters and $\Omega_\mathrm{m}$. This would be expected to lead to a degradation of the constraints obtained on $f$, but we leave a fuller investigation to subsequent work.

\section*{Acknowledgements}

We thank Andreu Font-Ribera for useful discussions. SN acknowledges funding from the Marie Sk\l odowska-Curie Actions under the H2020 Framework of the European Commission, project 660053 {\small COSMOVOID}, and a Dennis Sciama fellowship from the ICG. PC and WJP acknowledge support from the European Research Council through the Darksurvey grant 614030. The BigMultiDark simulations were performed on the SuperMUC supercomputer at the LeibnizRechenzentrum in Munich, using resources awarded to PRACE project number 2012060963. This work made use of the UK {\sm SCIAMA} High Performance Computing cluster supported by the ICG, SEPNet and the University of Portsmouth.



\bibliographystyle{mnras}
\bibliography{refs} 
%

\appendix

\section{Fitting an incomplete model using redshift space voids}
\label{sec:appendix}

The reconstruction method introduced in this paper in order to approximately recover real space void positions from redshift space galaxy data is more involved and computationally expensive than simply applying a void finder directly in redshift space, as been performed in numerous studies \citep{Paz:2013,Hamaus:2016,Hawken:2017,Hamaus:2017a,Achitouv:2017a}. We have argued in Section \ref{sec:complications} that the use of such redshift space voids is inconsistent with the assumptions made in the derivation of \emph{all} theory models of the void-galaxy correlation that have thus far been published in the literature, as none of these has included a model for the void-finding in the theory derivation. We also showed in Section \ref{sec:reconstruction} that using redshift space voids without reconstruction leads to large discrepancies between observed correlation multipoles and the predictions of the dispersion model of Eq. \ref{eq:xis dispersion}.

\begin{figure}
\begin{center}
\includegraphics[scale=0.5]{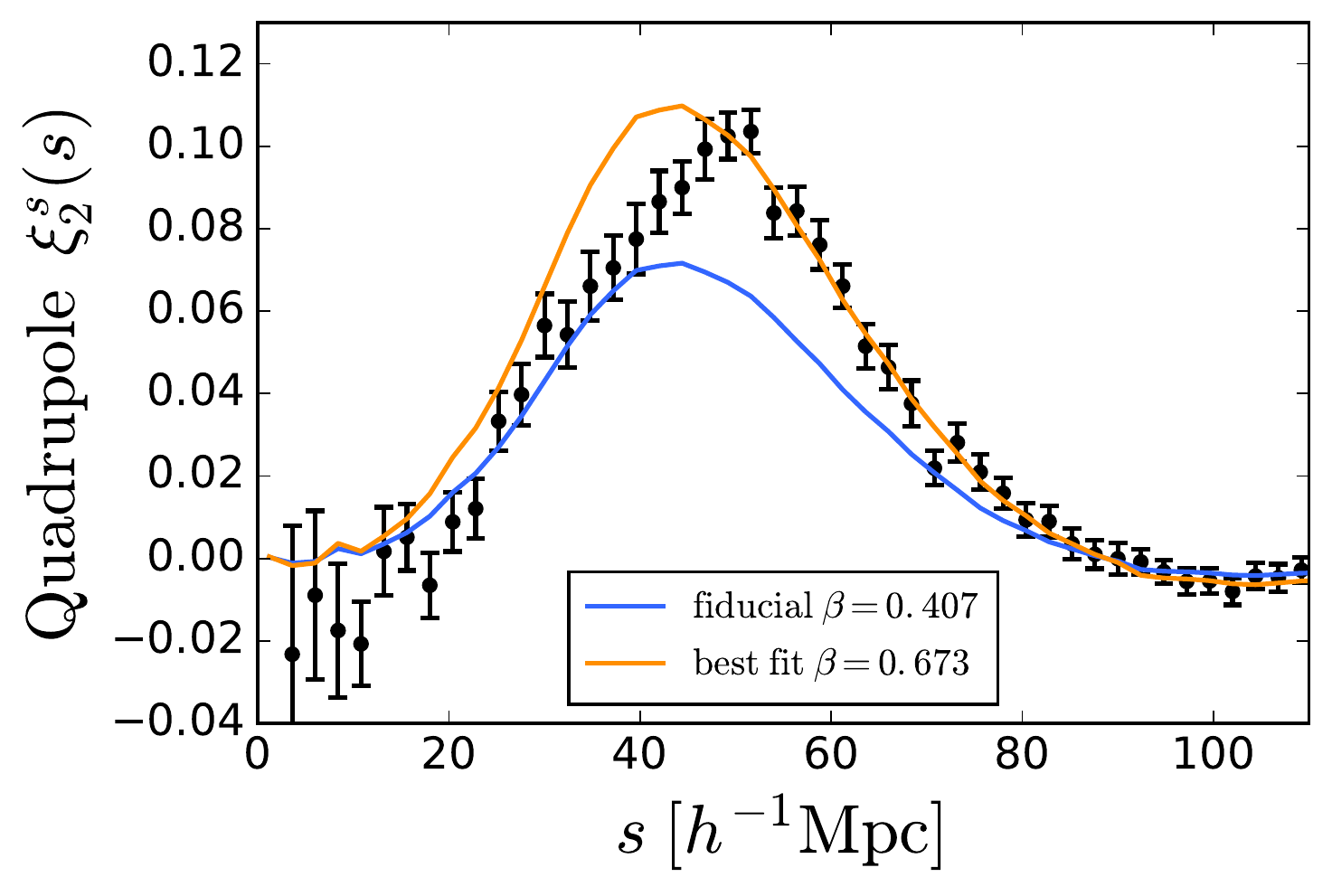}
\caption{The measured redshift space quadrupole $\xi^s_2(s)$ for voids in redshift space (data points with error bars) together with the theory predictions for the alternative RSD model of (\citealt{Cai:2016a,Hamaus:2017a}) for different values of $\beta\equiv f/b$. To enable a like-for-like comparison with \citealt{Hamaus:2017a}, for this figure void centres are taken to be the barycentre positions (see text).}
\label{fig:appendix}
\end{center}
\end{figure}

Nevertheless, the possibility remains that the use of an earlier theoretical model that does not include all of the terms argued as important by \citet{Nadathur:2018a} might by coincidence match the observed void-galaxy correlation for redshift space voids. In this Appendix, we consider the fit of the model of \citet{Cai:2016a} -- that has recently been used in an analysis of BOSS data by \citet{Hamaus:2017a} -- to redshift-space voids, to demonstrate that it does not match the data at the level required to make robust measurements.

\citet{Cai:2016a} derived expressions
\begin{equation}
\label{eq:Cai monopole}
\xi^s_0(s) = \left(1+\frac{\beta}{3}\right)\xi^r_0(s)\,,
\end{equation}
and
\begin{equation}
\label{eq:Cai quadrupole}
\xi^s_2(s) = \frac{2\beta}{3}\left[\xi^r_0(s)-\overline{\xi^r_0}(s)\right]\,,
\end{equation}
for the redshift space monopole and quadrupole respectively, where $\beta\equiv f/b$ and $\overline{\xi^r_0}(s)=3/r^3\int_0^r \xi^r_0(y)y^2 dy$. This model has the advantage that it allows measurement of $\beta$ through an estimator based on the redshift space quadrupole-to-monopole ratio, without requiring knowledge of the real space correlation. It is based on the same fundamental assumptions discussed in Section \ref{sec:complications} so should apply to the case where real space void positions are known, but comparison with simulation data in this case shows strong discrepancies within the void interior regions, $s<\overline{R_\mathrm{v}}$ \citep{Cai:2016a,Nadathur:2018a}. This also means that even if this model were to match data for the redshift space selected voids, this could only be due to coincidence, as the model assumptions are not satisfied. 

In Figure \ref{fig:appendix} we show the model prediction of Eq. \ref{eq:Cai quadrupole} for the fiducial value $\beta=0.407$ for our simulation and mock galaxy sample. The data points in the figure show the measured quadrupole for the correlation of voids defined directly in redshift space with the redshift space galaxy field. To enable a direct like-for-like comparison with the results of \citet{Hamaus:2017a}, we define the void centre positions to be the location of the Voronoi-volume-weighted barycentre of void member galaxy positions. (This results in slight differences relative to the $\xi^s_2(s)$ shown for the same redshift voids in Figure \ref{fig:fiducial_recon}, but does not affect the conclusions.) The covariance matrix has been re-estimated for this centre definition as well.

It can be seen that the measured and predicted quadrupoles share some qualitative features, in particular the absence of the strong negative `dip' feature seen for real space and reconstructed voids (Figure \ref{fig:fiducial_recon}). However, the fiducial model provides an extremely poor quantitative fit to the data: the reduced $\chi^2$ for the fit is $\sim9$. This is as expected given the discussion in Section \ref{sec:complications}: the assumptions on which this model is based are not valid for redshift space voids, therefore any similarities can only be due to coincidence.

If the poor $\chi^2$ value is ignored and a fitting procedure for $\beta$ is carried out, the resultant best-fit value is $\beta=0.673\pm0.016$, which is very strongly biased relative to the fiducial $\beta=0.407$. It is therefore clear that this method \emph{cannot} be used to self-consistently measure the growth rate.

It should be noted that the volume of our simulation box, $(2.5\;h^{-1}\mathrm{Gpc})^3$, is much larger than the available BOSS survey volume, and that the errors on our measurements of the correlation function multipoles are consequently much smaller than those currently achievable with survey data. Figure \ref{fig:appendix} suggests that given the much larger BOSS error bars, and the much wider radial binning used by \citet{Hamaus:2017a}, apparent consistency between the measured $\xi^s_2$ and this model is possible by coincidence. However, the much more precise measurements that will be possible with future survey data from DESI and Euclid will necessitate the use of the self-consistent reconstruction procedure.


\bsp	
\label{lastpage}
\end{document}